# The anisotropic Beer-Lambert law in $\beta - Ga_2O_3$: Spectral and polarization dependent absorption and photoresponsivity

Md Mohsinur Rahman Adnan[1], Darpan Verma[2], Chris Sturm[3], Mathias Schubert[4,5], and Roberto C. Myers[1,2*]

*[1) Department of Electrical and Computer Engineering, The Ohio State University, Columbus OH 43210, USA*

*[2) Department of Materials Science Engineering, The Ohio State University, Columbus OH 43210, USA*

*[3) Felix-Bloch-Institut für Festkörperphysik, Universität Leipzig, Linnéstr. 5, 04103 Leipzig, Germany*

*[4) Department of Electrical and Computer Engineering, University of Nebraska-Lincoln, Lincoln NE 68588, USA*

*[5) NanoLund and Solid State Physics, Lund University, S-22100 Lund, Sweden*

**Due to its low symmetry, $\beta - Ga_2O_3$ exhibits a strongly anisotropic optical response. As a result, the absorption spectra change with the polarization state of the incoming photons. To understand this phenomenon, here we calculate the complete electromagnetic wave equation solutions as a function of linear polarization angle and photon energy for $\beta - Ga_2O_3$ using its previously measured complex dielectric function tensor. The significant off-diagonal terms in this tensor can result in a non-exponential decay in the photon flux, indicating that the Beer-Lambert law is not generally valid in this anisotropic material. However, for above-band-gap spectral regions which depend on crystallographic orientations ($> 5.8\,eV$ (001 plane), $> 5.2\,eV$ (010 plane)) an effective absorption coefficient well approximates the photon flux decay with depth. On the other hand, near the optical absorption edge ($4.9 - 5.8\,eV$ (001 plane), $4.65 - 5.2\,eV$ (010 plane)) the photon flux decay exhibits a sum of two exponential decays, such that two effective absorption coefficients are necessary to model the loss behavior versus the absorption depth. This behavior manifests from the presence of dichroism in $\beta - Ga_2O_3$. A single effective absorption coefficient can only be recovered for this energy range by augmenting the isotropic Beer-Lambert law with a critical penetration depth and polarization dependence. Using these results, we calculate the polarization-dependent photoresponsivity spectra for light polarized along different crystallographic directions.**

## INTRODUCTION

Monoclinic $\beta - Ga_2O_3$ can support high breakdown fields in Radio Frequency (RF) and power electronic applications thanks to its ultra-wide bandgap (UWBG) [1–4]. Moreover, $\beta - Ga_2O_3$ deep ultra-violet (UV) photodiodes have been demonstrated with large responsivities [5–10]. As reported by room temperature transmittance and reflectance spectra measurements, the optical absorption edge ($E_{edge}$) in $\beta - Ga_2O_3$ for light polarized along the a-axis [100] and c-axis [001] almost coincide in the range $\sim 4.48 - 4.57\,eV$, but for light polarized along b-axis [010] $E_{edge}$ is at a much higher energy $\sim 4.73-5.06\ eV$ [1,2,11]. This happens due to the optical selection rules governing the allowed interband optical transitions combined with the crystal field splitting of the valence bands [3,11,12]. The Ga 4s state's conduction band minimum (CBM) is isotropic with $\Gamma_1^+$ as its irreducible representation. The valance band maximum (VBM) at the Γ point consists of O 2p states. These states produce bands with $\Gamma_1^+, \Gamma_2^+, \Gamma_1^-$ and $\Gamma_2^-$ irreducible representations. Only VBM $\Gamma_1^-$ to CBM $\Gamma_1^+$ is dipole allowed for light polarized along b-axis, while VBM $\Gamma_2^-$ to CBM $\Gamma_1^+$ is dipole allowed for light polarized in the a-c plane (010) orientation. This results in a blueshift of $E_{edge}$ by $\sim 0.35\ eV$ from a-c plane polarized to b-axis polarized light [1,13]. Using this polarization anisotropy between b- and c-axes, b-c plane (100) oriented $\beta - Ga_2O_3$ has recently been used to design a highly narrow-band ($\sim 0.23\ eV$) solar-blind photodetector ($4.52\ eV < E_{ph} < 4.76\ eV$) [10]. Since the $E_{edge}$ order is $b > a > c$ ($4.90\ eV > 4.56\ eV > 4.54\ eV$), a more selective deep-UV solar-blind photodetector is expected for a-b plane in a (001) oriented crystal [1,10]. This inspired us to reveal the polarization dependent anisotropic absorption behavior and photoresponsivity spectra of (001) oriented $\beta - Ga_2O_3$. We further analyze (010) oriented $\beta - Ga_2O_3$ to obtain the anisotropic characteristics in the a-c plane and compare the results with those from the a-b plane.

Although it is well-known that $\beta - Ga_2O_3$ displays highly anisotropic optical absorption, a quantitative model for the anisotropic optoelectronic processes taking place in this material is not yet well-established. In particular, in the broad semiconductor optoelectronic field (including photovoltaics), a fundamental assumption is the applicability of the Beer-Lambert (BL) law, which states that the photon flux ($\Phi$) decays exponentially with penetration depth (z) as it transmits into the absorbing medium, $\Phi = \Phi_0 e^{-\alpha z}$, where $\Phi_0$ is the photon flux at z = 0, and the exponent $\alpha$ is the material's absorption coefficient [14]. In this fundamental paradigm, the generation rate of photocarriers is $G = -\frac{d\Phi}{dz} = \Phi_0 \alpha e^{-\alpha z}$. The photocurrent is then estimated by considering the photocarrier collection processes (electron-hole drift-diffusion) and competing recombination rates.

However, as will be shown in our work here, the BL law in this formulation is only valid for isotropic materials. Specifically, the definition of the BL law requires the existence of a plane wave electromagnetic eigenmode whose polarization state does not alter upon propagation along a certain distance. This is trivially given for isotropic materials, and such eigenmodes also exist for all anisotropic materials whose major polarizability axes are orthogonal to each other, for example, in crystal structures with hexagonal, trigonal, tetragonal, and orthorhombic lattice systems. For highly anisotropic materials such as monoclinic symmetry $\beta - Ga_2O_3$, α cannot be defined immediately and the derivation of the generation rate and photocurrent generation process must be re-examined.

We establish here that the optical phenomenon known as dichroism is present in $\beta - Ga_2O_3$. It is also known that $\beta - Ga_2O_3$ exhibits birefringence property [15,16]. Together these physical properties give rise to the deviation from the isotropic BL law. We find that in deviation from the traditional formulation of the fundamental BL law, one must introduce wavelength, penetration depth, and polarization dependent absorption coefficients to approximate the otherwise complex photo absorption behavior in $\beta - Ga_2O_3$. This deviation finds its explanation in the fact that eigenmodes whose polarization state is unaltered upon propagation distances do not generally exist in materials with low symmetry such as in monoclinic symmetry $\beta - Ga_2O_3$.

In the monoclinic $\beta - Ga_2O_3$ there is no possibility to orient the crystal above the bandgap such that the dielectric tensor is purely diagonal because of finite imaginary parts that remain in the off-diagonal components. Hence, the isotropic BL law must indeed be rewritten. Here we attempt introducing an effective isotropic absorption coefficient in the BL law with an augmented penetration depth dependence, energy dependence, and polarization dependence to maintain the appearance of the isotropic BL law characteristic, and to approximate the complex absorption behavior in the highly anisotropic $\beta - Ga_2O_3$. Once we obtain the flux decay behavior, we calculate the anisotropic photoresponsivity spectra for light polarized along different crystallographic directions in $\beta - Ga_2O_3$.

## I. OPTICAL ABSORPTION AND PHOTOCURRENT MODEL

Recent theoretical and experimental work demonstrates that the dielectric function of $\beta - Ga_2O_3$ contains large off-diagonal tensor components in contrast to cubic or hexagonal semiconductors such as $GaN, InN, AlN$ and their alloys which are used in current optoelectronic applications [17,18]. These off-diagonal dielectric tensor components of $\beta - Ga_2O_3$ cannot be orthogonalized by simple choice of coordinate rotations [19]. Here we utilize the measured complex dielectric function tensor ($\boldsymbol{\varepsilon}$) of $\beta - Ga_2O_3$ by Sturm et al. [17], to numerically model the anisotropic Electro-Magnetic (EM) wave propagation and decay within $\beta - Ga_2O_3$. The internal transmitting time-averaged Poynting vector ($\overrightarrow{S_t}$) is calculated as a function of photon energy ($E_{ph}$), penetration depth ($z$), and polarization angle (θ) of the incident EM wave from which the photon flux, $\Phi = \frac{|\overrightarrow{S_t}|}{E_{ph}}$, within $\beta - Ga_2O_3$ is found. The $E_{ph}$, θ, and z dependent photocarrier generation rate can then be numerically calculated from the full EM solutions,

$$G = -\frac{d\Phi}{dz} = -\frac{d|\overrightarrow{S_t}|}{E_{ph}\,dz} = -\frac{d\left|\frac{1}{4}\left(\overrightarrow{E_t} \times \overrightarrow{H_t^*} + \overrightarrow{E_t^*} \times \overrightarrow{H_t}\right)\right|}{E_{ph}\,dz}\,, \tag{1}$$

where $\overrightarrow{E_t}(\overrightarrow{E_t^*})$ and $\overrightarrow{H_t}(\overrightarrow{H_t^*})$ are the complex electric and magnetic field components (conjugates of components) of the EM wave solutions, respectively, within the medium with $\boldsymbol{\varepsilon}$. Knowing the incoming power density, i.e., the time-averaged Poynting vector magnitude at z = 0 before entering the medium, $|\overrightarrow{S_{in}}|$, the photoresponsivity spectra due to photon absorption in the medium, $I_{PR}$ can be calculated as a function of polarization angle θ integrating the numerically calculated generation rate $G$ over a region from the entrance interface with air at $z = 0$ to a depth at $z = d$,

$$I_{PR} = \frac{q \int_0^d G dz}{|\overrightarrow{S_{in}}|}, \quad (2)$$

where q is the charge per generated photocarrier. In this initial approach, we ignore any $z$ dependence of the photocarrier collection processes to isolate the effect of the polarization dependence due to the anisotropic dielectric function. Eq. (2) could be later modified to include $z$ and bias dependent quantum efficiency that are expected to occur due to exciton dissociation and minority carrier diffusion limits. The $I_{PR}$ spectra in a $\beta - Ga_2O_3$ region of thickness $z$ as a function of θ are calculated in this work, which explain the multiple anisotropic excitonic peaks we previously reported [20].

The polarization-dependent exciton absorption in $\beta - Ga_2O_3$, has been demonstrated both theoretically, by first principles calculations, and experimentally, by spectroscopic ellipsometry [13,17,18,21–23]. Note that in Ref. [20] we assigned the principal axes labels of the of the eXcitons (previously $X_c \sim 4.65\ eV, X_a \sim 4.9\ eV, X_{ac} \sim 5 - 5.2\ eV$ and $X_b \sim 5.5\ eV$) based on the DFT study in Ref. [22]. Mock et al. clarified the transition dipole orientation of these transitions [23] through combined DFT and spectroscopic ellipsometry as $X_c \sim 4.9\ eV$, $X_a \sim 5.2\ eV$, and $X_b \sim 5.5\ eV$, which we now adopt. It is worth noting that while the $X_b$ dipole is purely along the b-axis, neither the $X_c$ nor $X_a$ transitions are purely along either c-axis or a-axis, but lie at some angle within the a-c plane [23]. The numerical $\Phi$ data are examined as a function of $E_{ph}$, θ, and $z$ to test the range of applicability of the isotropic BL law in $\beta - Ga_2O_3$. We find a significant deviation from the isotropic BL law near $E_{edge}$ range of $\beta - Ga_2O_3$ where the excitonic peaks appear [17,22,23]. However, an exponential decay model with two absorption coefficients can largely account for the $\Phi$ decay with $z$ and may be adequate for predicting the anisotropic absorption process in this energy range for $\beta - Ga_2O_3$. In special cases, such as when the excitation is polarized along a major axis, the isotropic BL law is recovered, but in unpolarized or elliptically polarized conditions, our results show that a full EM solution-based model is necessary to fully predict the absorption, reflection, and transmission characteristics of $\beta - Ga_2O_3$, and, by extension, its family of monoclinic alloys [24].

To acquire the transmitted power density, $|\overrightarrow{S_t}|$, we solve the EM wave equation for light incident on an Air/ $\beta - Ga_2O_3$ interface and travelling along +z into $\beta - Ga_2O_3$ using the Berreman-Schubert theory [25,26]. The directions of the Poynting Vectors incident upon ($\overrightarrow{S_{in}}$), reflected from ($\overrightarrow{S_r}$) and transmitted into ($\overrightarrow{S_t}$) the $\beta - Ga_2O_3$ sample with respect to the lab frame are shown in Figure 1(a). Figures 1(b) and 1(c) show the different directions including the major axes [100], [010] and [001] of the $\beta - Ga_2O_3$ sample given $\overrightarrow{S_t}$ is travelling perpendicular to the (001) and (010) planes, respectively. Full details of the model are provided in the Appendix. Finally, Figures 1(d) and 1(e) show the polarization rotation of light incident upon (001) and

(010) planes at steps of $\Delta\theta = 15^0$ and $\Delta\theta = 30^0$ respectively. A few directions of interest, e.g., excitonic transition dipole moment orientations $X_b$ $(\theta = 90^0)$ in the (001) plane, $X_c$ $(\theta \cong 115.1^0)$, and $X_a(\theta \cong 25.2^0)$ in the (010) plane [23], along with the major axes a- [100] $(\theta = 0^0)$ in both panes, b- [010] $(\theta = 90^0)$ in the (001) plane and c- [001] $(\theta = 103.8^0)$ in the (010) plane of $\beta - Ga_2O_3$, are also denoted in these figures. The color coding of Figs. 1(d) and 1(e) is maintained for other figures in this work.

## II. POLARIZATION-DEPENDENT REFLECTION, ABSORPTION, AND TRANSMISSION SPECTRA

For light incident on the interface, there is considerable reflection due to refractive index mismatch. Taking this into account, the reflectance, transmittance, and absorbance spectra at any $z$ or $\theta$ are, respectively, defined as, $R = \frac{|\overrightarrow{S_r}|}{|\overrightarrow{S_{in}}|}, T = \frac{|\overrightarrow{S_t}|}{|\overrightarrow{S_{in}}|}, A = \frac{|\overrightarrow{S_{in}}| - |\overrightarrow{S_r}| - |\overrightarrow{S_t}|}{|\overrightarrow{S_{in}}|}$. The conservation of energy principle then is observed at any depth $z$ as, $R + T + A = 1$. Within $\beta - Ga_2O_3$, $R = 0$ and $A = 1 - T$. It should be noted that while calculating the values of $R$, $T$ and $A$, no polarization discrimination was applied, i.e., these terms contain all reflected, transmitted, and absorbed EM waves regardless of them having multiple eigenmodes. For example, the calculation of $R$ contains all s- and p-polarized reflected EM waves; the same is true for $T$ and $A$.

The interface reflectance spectra $R$ for light incident at different polarization angles θ from Air/ (001) $\beta - Ga_2O_3$ and Air/ (010) $\beta - Ga_2O_3$ interfaces are presented in Figures 2(a) and 2(d), respectively. As expected, multiple peak features can be seen in the spectra in the near $E_{edge}$ energy range. Variation in light polarization is observed to modulate the $R$ spectra by varying (increasing/decreasing) the different peak amplitudes, but never shifting their positions. Similar variation of amplitude of multiple peaks has been observed in previous anisotropic excitonic photocurrent measurements in $\beta - Ga_2O_3$ [20]. This observation confirms that these peak features have their origin in the anisotropic nature of $\beta - Ga_2O_3$ absorption behavior modulated by multiple excitonic transitions. The excitonic transition peaks [23] i.e., $X_c$, $X_a$, and $X_b$ are shown by reference lines in Fig. 2. Indeed, the first pronounced peak in $R$ spectra from Air/ (001) $\beta - Ga_2O_3$ interface appear at the peak energy $X_a$ as can be seen in Figure 2(a). The $R$ spectra from Air/ (010) $\beta - Ga_2O_3$ interface show two pronounced peaks near $E_{edge}$, one at $X_c$ and another at $X_a$ energies, as seen in Fig. 2(d). The absorbance and transmittance spectra into $\beta - Ga_2O_3$ below the surface at $z = 1\,nm$ is calculated as $A(z = 1\,nm) = A_0 = \frac{|\overrightarrow{S_{in}}| - |\overrightarrow{S_r}| - |\overrightarrow{S_{t0}}|}{|\overrightarrow{S_{in}}|}$, $T(z = 1\,nm) = T_0 = \frac{|\overrightarrow{S_{t0}}|}{|\overrightarrow{S_{in}}|}$. While $A_0$ gives the amount of light absorbed within a small penetration depth $(z = 1\,nm)$ into $\beta - Ga_2O_3$, $T_0$ gives the amount of light that remains unabsorbed after passing through this depth. $A_0$ is shown for (001) and (010) $\beta - Ga_2O_3$ in Figures 2(b) and 2(e). While $A_0$ starts to rise near the $X_c$ transition for the (001) plane, this rise starts well before the $X_c$ transition for the (010) plane. This observation points towards $E_{edge}$ anisotropy in $\beta - Ga_2O_3$. $T_0$ for the (001) and (010) planes are shown in Figures 2(c) and 2(f), respectively. Just as $R$ spectra, $T_0$ spectra are also strongly modulated by the excitonic transitions. There are pronounced dips in $T_0$ spectra at the excitonic transition peaks, pointing to the fact that the presence of excitons reduces flux transmission into $\beta - Ga_2O_3$. This is expected since the excitonic transitions facilitate absorption of photons in $\beta - Ga_2O_3$.

## III. PENETRATION DEPTH-DEPENDENT PHOTON FLUX

The flux $\Phi$ is directly proportional to the transmittance factor $T$ by its definition. We define the incoming photon flux into $\beta - Ga_2O_3$ inside the surface at $z = 1\,nm$ as, $\Phi_0 = \frac{|\overrightarrow{S_{t0}}|}{E_{ph}} = \frac{T_0}{E_{ph}}|\overrightarrow{S_{in}}|$. At any larger depth $z$ inside $\beta - Ga_2O_3$, the photon flux is given as, $\Phi = \frac{|\overrightarrow{S_t}|}{E_{ph}} = \frac{T}{E_{ph}}|\overrightarrow{S_{in}}|$. The factor $\frac{\Phi}{\Phi_0} = \frac{|\overrightarrow{S_t}|}{|\overrightarrow{S_{t0}}|} = \frac{T}{T_0}$ gives the decay of photon flux within $\beta - Ga_2O_3$. Knowing all these terms the flux decay, $\frac{\Phi}{\Phi_0}$ can be obtained.

The use of the isotropic BL law to predict $\frac{\Phi}{\Phi_0}$ to obtain a generalized absorption coefficient, $\alpha$ was not possible for $E_{ph} \leq 4.90\,eV$ in the (001) plane and for $E_{ph} \leq 4.65\,eV$ in the (010) plane. This is because the ellipsometry data reported in the literature [17,18,23] are collected using specialized instruments which possess only limited sensitivity to small components of the imaginary parts of the anisotropic dielectric functions for photon energies below the exciton energies. As discussed by Fujiwara [27], noise in absorption coefficients evaluated from ellipsometry measurements appears when $\alpha < 10^3\,cm^{-1}$. Depending on the real part of the complex-valued index of refraction this translates to sensitivity limits in the extinction coefficients of around 0.1 to 0.01. The reported data for the imaginary parts of the dielectric function tensor spectra for $\beta - Ga_2O_3$, due to noise level of the measuring instrument, then can only be used reliably for photon energies from shortly below the exciton energies onwards to the higher photon energy limit of 8.65 eV. Given these considerations we only focus on the applicability of the isotropic BL law to predict the flux decay above $E_{edge} \sim 4.5\,eV$ in our work. The choice of this limit is supported by Sturm et al.'s work [17] where the imaginary parts of the measured dielectric tensor were described to have non-zero values predicting finite absorption for photon energies above 4 $eV$.

The photon flux decay behavior in the photon energy ranges $E_{edge} < E_{ph} \leq E_{ph}^0$, given $E_{ph}^0 = 4.9\,eV$ in the (001) plane and $= 4.65\,eV$ in the (010) plane, must be calculated numerically from the EM wave equation solution and cannot be estimated in confidence with a generalized absorption coefficient, $\alpha$. Thus, the isotropic BL law is not normally applicable for $E_{edge} < E_{ph} \leq E_{ph}^0$ photons being transmitted-absorbed in $\beta - Ga_2O_3$ for most of the engineering applications including photodetectors, where a device active region of size of $z \leq 1\,\mu m$ is the norm [5–7,28]. Nevertheless, we can still obtain an estimate of $\alpha$ for the energy range $4.5\,eV \leq E_{ph} \leq 4.9\,eV$ in the (001) plane by best-matching the $\frac{\Phi}{\Phi_0}$ versus $z$ behavior using the isotropic BL law within $z \leq 1\,\mu m$ depth from the air/ $\beta - Ga_2O_3$ interface; here the choice of $z$ is energy dependent. Similarly, an estimate of $\alpha$ in the (010) plane for the range $4.5\,eV \leq E_{ph} \leq 4.65\,eV$ can be obtained.

Now we look at the transmission of photons in (001) $\beta - Ga_2O_3$ with energies above $4.9\,eV$. Results for only a few polarization directions incident onto this plane are presented in Figure 3, although the conclusions reached here are valid for other polarization directions as well. We fit the decay behavior for at least three orders of magnitude change ($\Delta(\frac{\Phi}{\Phi_0}) \geq 10^3$) to reach our conclusion in all cases. The $\frac{\Phi}{\Phi_0}$ versus $z$ data up to $z = 10\,\mu m$ in the range $4.9\,eV < E_{ph} \leq 5.8\,eV$ and for polarization directions [100] ($\theta = 0^0$), [140] ($\theta \cong 45^0$) and [010] ($\theta = 90^0$) in the (001) plane are shown in Figures 3(a)-3(c), respectively. The $\frac{\Phi}{\Phi_0}$ versus $z$ plots for polarized light along the major axes in Figures 3(a) and 3(c) are shown along with their isotropic BL law fits (solid lines) going through the data (points). The non-major axes polarized light transmission case is shown in Figure 3(b) with $\frac{\Phi}{\Phi_0}$ versus $z$ data points for direction [140]. The data points can be fitted with a two-absorption coefficient exponential decay model as shown by the dashed lines. This behavior is found consistently true for all non-major axes' directions between the major axes in the (001) plane. The origin of such behavior is discussed later in our work. Beyond $5.8\,eV$, for energies in the range $5.8\,eV < E_{ph} \leq 8.65\,eV$, the $\frac{\Phi}{\Phi_0}$ versus $z$ data points for polarization directions [100], [140] and [010] at steps $\Delta E_{ph} = 0.3\,eV$ are shown in Figures 3(d)-3(f), respectively. For all these cases, the data were fitted up to $z = 0.5\,\mu m$ with isotropic BL law expression denoted by solid line fits. These results show that for photon energies above $5.8\,eV$, the isotropic BL law can predict the flux decay in (001) $\beta - Ga_2O_3$ for any polarization angle $\theta$.

Next, the decay of photon flux in (010) $\beta - Ga_2O_3$ above $4.65\,eV$ is examined. The transmission of photons with different polarization angles $\theta$ and energies in the range $4.65\,eV < E_{ph} \leq 5.2\,eV$ is confirmed to be follow the decay behavior described by a two absorption coefficient exponential decay model as was found for non-major axes direction in the (001) plane. Examples are shown with $\frac{\Phi}{\Phi_0}$ versus $z$ data points and their dashed line fits up to $z = 10\,\mu m$ for sample directions [100] ($\theta = 0^0$), [102] ($\theta \cong 50^0$) and [001] ($\theta = 103.8^0$) in the (010) plane in Figures 4(a)-4(c), respectively. Beyond $5.2\,eV$, for energies in the range $5.20\,eV < E_{ph} \leq 8.65\,eV$, the $\frac{\Phi}{\Phi_0}$ versus $z$ decay follows the isotropic BL law for any polarization angle $\theta$ in the (010) plane. The data points fitted with the isotropic BL law expression (solid lines) are shown at steps $\Delta E_{ph} = 0.3\,eV$ up to $z = 0.5\,\mu m$ for directions [100], [102], [001] in Figures 4(d)-4(f), respectively.

The (001) plane of $\beta - Ga_2O_3$ (c2mm) has two mirror planes that divide the total $360^0 = 4 \times 90^0$ into four equivalent quadrants. The decay behavior along the major axes a- [100] ($\theta = 0^0$) and b- [010] ($\theta = 90^0$) in (001) $\beta - Ga_2O_3$ follows the isotropic BL law above 4.9 $eV$ till 8.65 $eV$ (the highest limit of Sturm et al.'s calculated energies). For in-between polarization angles ($0^0 < \theta < 90^0$), there are two different energy ranges of interest we must examine to describe the photon flux decay behavior. For energies in the range $4.9\, eV < E_{ph} \leq 5.8\, eV$, the decay can be explained with a two-absorption coefficient exponential decay model of the form,

$$\frac{\Phi}{\Phi_0} = \begin{cases} e^{-\alpha}, & 0 < z \leq z_c \\ e^{-\alpha z_c} e^{-\alpha^*(z - z_c)}, & z_c < z \end{cases}, \tag{3}$$

where $z_c$ is a $E_{ph}$ and $\theta$ dependent critical thickness parameter within which the isotropic BL law holds true. For any depth beyond $z_c$, i.e., $z > z_c$, the exponential decay constant weakens from $\alpha$ to $\alpha^*$. On the other hand, above 5.8 $eV$, the isotropic BL law works well with only $\alpha$ to describe the depth dependent photon flux decay for any polarization angle $0^0 \leq \theta \leq 90^0$ in the (001) plane.

In contrast to the higher symmetry (001) projection discussed above, the (010) projection of $\beta - Ga_2O_3$ (p2) has only a two-fold rotational symmetry on the plane dividing the total $360^0 = 2 \times 180^0$ into two equivalent halves. In this plane, the expression in Eq. (3) describes the flux decay behavior for any polarization angle $0^0 \leq \theta \leq 180^0$ including the major axes a- [100] ($\theta = 0^0$) and c- [001] ($\theta = 103.8^0$) within the range $4.65\, eV < E_{ph} \leq 5.20\, eV$. Beyond 5.2 $eV$, the isotropic BL law with absorption coefficient $\alpha$ works well to describe the data for any polarization angle $0^0 \leq \theta \leq 180^0$ in the (010) plane.

The behavior of Eq. (3) arises from the fact that the two forward propagating eigenmodes of the incident wave are added together while computing the intensity of the transmitted wave. These eigenmodes receive field amplitudes which decay differently with penetration depth depending on the polarization state of the incoming wave. The difference in the field amplitude decay for the two forward propagating eigenmodes is a result of dichroism; specifically linear dichroism observed in $\beta - Ga_2O_3$. Linear dichroism is defined as the difference in the absorption of light incident at linear polarization directions that are perpendicular to each other. Linear dichroism manifests in $\beta - Ga_2O_3$, as shown in Figure 5. Figures 5(a) and 5(b) show the linear dichroism in the (001) and (010) plane, respectively. The absorbance is calculated as $A = \frac{|\overrightarrow{S_{in}}| - |\overrightarrow{S_r}| - |\overrightarrow{S_t}|}{|\overrightarrow{S_{in}}|}$ at any given depth $z$. Linear Dichroism (LD) in the (001) plane is defined as the difference in absorbance along the a-axis and the b-axis i.e., $LD_{a-b} = A_{a-axi} - A_{b-axi}$ since they are at $90^0$ to each other. For the (010) plane it is defined as the difference in absorbance along the $X_c$ dipole moment direction ($\theta \cong 115^0$ on the (010) plane) and the $X_a$ dipole moment direction ($\theta \cong 25^0$ on the (010) plane) i.e., $LD_{X_c - X_a} = A_{X_c\ at\ 115^0} - A_{X_a\ at\ 25^0}$ since they are at $90^0$ to each other. It is evident that LD for both planes increase as the penetration depth increases. The LD in the (001) plane is observed to be significant in the energy range $4.9\ eV < E_{ph} \leq 5.8\ eV,$ while in the (010) plane LD dominates in $4.65\ eV < E_{ph} \leq 5.2\ eV$ range. These are the same energy ranges over which the isotropic BL law with a single $\alpha$ fails to describe the photon flux decay behavior and requires the addition of a secondary weakened absorption coefficient $\alpha^*$ as in Eq. (3) to describe absorption in $\beta - Ga_2O_3$. Linear dichroism gives rise to a rotation of the polarization state of the incoming wave i.e., a rotation of the main field direction and change in phase continuously as the wave travels into $\beta - Ga_2O_3$. The physical reason is that the absorbers in the material – the band-to-band transitions and associated excitons – are not perpendicular to each other in their major linear absorption (dipole) directions. The nonlinear intensity versus penetration depth behavior for any given energy can be described employing a sum of two exponential decays with different absorption coefficients,

$$I_{tot}(P) = I_1(P)e^{-\alpha_1 z} + I_2(P)e^{-\alpha_2 z} \cong I_{eff}(P)e^{-\alpha_{eff}(P,z)z}, \quad (4)$$

The $(P)$ in Eq. (4) denotes the polarization state of the incident wave. Eq. (4)'s equivalence is described in Eq. (3) by introducing a piecewise constant penetration depth dependence defined as the critical thickness, $z_c$. By approximating the nonlinear intensity versus penetration depth variation, we regain the BL law in a simple isotropic form to describe the complex anisotropic absorption behavior of $\beta - Ga_2O_3$; but at the same time, we add the complexity of considering the wavelength, polarization state, and penetration depth of the incident wave when calculating the absorption coefficient of $\beta - Ga_2O_3$.

## IV. POLARIZATION-DEPENDENT ABSORPTION COEFFICIENT

The BL law absorption coefficient $\alpha$ for photons with energy $4.9\ eV < E_{ph} \leq 8.65\ eV$ and different polarization angles $0^0 \leq \theta \leq 90^0$ is plotted in solid lines in Figure 6(a). It must be noted that in the energy range $4.9\ eV < E_{ph} \leq 5.8\ eV$ and for polarization angles $0^0 < \theta < 75^0$, the absorption coefficient $\alpha$ is applicable only to a limited depth of $z < z_c$. Beyond the critical penetration depth $z > z_c$, Eq. (3) modifications will describe the decay behavior in this near $E_{edge}$ energy range and $\alpha^*$ is needed. Beyond $5.8\ eV$, $\alpha$ alone is sufficient for the angles $0^0 < \theta < 75^0$. The $\alpha^*$ is also plotted in Figure 6(a) in dots for the energy range $4.9\ eV < E_{ph} \leq 5.8\ eV$ and for the sampled polarization angles $0^0 < \theta < 75^0$. Interestingly, they all approach the lowest limit of $\alpha$ i.e., the b-axis [010] direction values in the (001) plane. This b-axis is the direction of $X_b$ exciton's dipole moment orientation which shows an obvious modulation of the absorption behavior by the excitonic transition. Finally, for angles $75^0 < \theta < 90^0$ in the (001) plane, $\alpha$ alone is sufficient for the energy range $4.9\ eV < E_{ph} \leq 8.65\ eV$. We also show the less reliable estimates of $\alpha$ in the energy range $4.5\ eV \leq E_{ph} \leq 4.9\ eV$ for completeness. The inset shows the same spectra on a semi-log scale in the range $4.5\ eV \leq E_{ph} \leq 6.5\ eV$ to clearly show the anisotropy of $\alpha$. From the inset, the $\alpha$ along b-axis drops to a value as small as $\sim 10^2/cm$ at $\sim 4.9\ eV$ compared to a-axis which is still at $\sim 10^3/cm$ even at $\sim 4.5\ eV$. This points to the origin of the $E_{edge}$ anisotropy between a- and b-axes in the (001) plane.

The BL law absorption coefficient $\alpha$ for photons with energy $4.65\ eV < E_{ph} \leq 8.65\ eV$ and different polarization angles $0^0 \leq \theta \leq 180^0$ is plotted in Figure 6(b). In the energy range $4.65\ eV < E_{ph} \leq 5.2\ eV$ and for polarization angles $0^0 \leq \theta \leq 140^0$, the absorption coefficient $\alpha$ is applicable only to a limited depth of $z < z_c$, beyond which $\alpha^*$ is needed. Above $5.2\ eV$, $\alpha$ can describe the decay sufficiently for the angles $0^0 \leq \theta \leq 140^0$. The $\alpha^*$ for sampled polarization angles in range $0^0 \leq \theta \leq 140^0$ for energy range $4.65\ eV < E_{ph} \leq 5.2\ eV$ are plotted in dots in Figure 6(b). They all approach the lowest limit of $\alpha$ i.e., the $\theta \cong 25^0$ values in the (010) plane. This $\theta \cong 25^0$ is the direction of $X_a$ exciton's dipole moment orientation which shows an obvious modulation of the absorption behavior by the excitonic transition. Finally for angles $140^0 < \theta < 180^0$ in the (010) plane, $\alpha$ alone is sufficient for the energy range $4.65\ eV < E_{ph} \leq 8.65\ eV$. The less reliable estimates of $\alpha$ in the energy range $4.5\ eV \leq E_{ph} \leq 4.65\ eV$ are included for completeness. The inset highlights the anisotropy of the absorption edge plotted over the same range as in Fig. 6(a) for comparison. The $\alpha$ along a-axis drops to a value as small as $\sim 10^3/cm$ in the same energy range (within $\sim 0.1\ eV$ of $E_{edge} \sim 4.5\ eV$) as the c-axis, which is a far milder anisotropy than in the (001) plane. A clearer anisotropy between the two major axes can only be seen at a higher $\alpha \sim 10^4/cm$ and the $\alpha$ rises to this value at a higher energy for light polarized along a-axis compared to c-axis. Thus, the $E_{edge}$ ordering $b > a > c$ is established from our calculations.

In (001) $\beta - Ga_2O_3$ the critical thickness $z_c$ appears to be polarization and energy dependent; its smallest value appearing at around $E_{ph} \sim 5.3\ eV$ as shown in Figure 7(a). This is the energy at which $A_0$ spectra for the two perpendicular directions, a-axis and b-axis, are separated the most on the (001) plane i.e., the LD just below the surface is greatest. The $z_c$ at any given energy drops to lower values with increasing polarization angle i.e., as the light polarization rotates from a- axis towards b-axis in the (001) plane. In (010) $\beta - Ga_2O_3$, the smallest value of $z_c$ appears at around $E_{ph} \sim 5.05\ eV$ as shown in Figure 7(b). This again is the energy at which $A_0$ spectra for the two perpendicular directions aligned with $X_a$ and $X_c$ dipole orientations are separated the most on the (010) plane i.e., the LD just below the surface is greatest. The $z_c$ for any given energy has its minimum value near the $X_a$ dipole orientation direction which increases towards the $X_c$ dipole orientation direction in the (010) plane. These observations demonstrate an unusual deviation from the historic isotropic BL law absorption process within (001) $\beta - Ga_2O_3$. The $z_c$ dependence of the photon flux decay should always be considered to properly incorporate the effects of near $E_{edge}$ excitonic transitions when describing the anisotropic absorption process in

$\beta - Ga_2O_3$.

To summarize the discussion above, for incoming light of any polarization direction, the absorption coefficient $\alpha$ weakens to a value of $\alpha^*$ that is equal to the absorption coefficient $\alpha$ of least absorbing polarization direction on the plane, i.e., the b-axis on the (001) plane or the $X_a$ dipole orientation direction on the (010) plane, beyond a critical penetration depth $z_c$. The critical penetration depth $z_c$ itself is dependent on the polarization of light incident on the slab, and takes the minimum value, i.e., needs less amount of penetration when the incident light is aligned closest to this least absorbing polarization direction. This effect arises from linear dichroism in $\beta - Ga_2O_3$. For a material exhibiting linear dichroism, the incoming polarization state of the light will be rotated upon transmission. If it is an ideal linear polarizer, usually the differential absorption will have a very high value and the incoming polarization state will rotate almost completely to the state of least absorption [29]. In the case of $\beta - Ga_2O_3$, strong linear dichroism effectively rotates the incoming linear polarization state to that of the least absorption; and it takes less penetration to reach the state of least absorption if the initial state was already closely aligned. It should be noted that, since there is birefringence present in $\beta - Ga_2O_3$, ellipticity develops as light travels deeper into the material. The rotation of the polarization state, both linear and circular, needs further investigation to understand the complete mechanism of light travelling into $\beta - Ga_2O_3$.

Figure 8 shows polar plots of the full angular dependence $0^0 \leq \theta \leq 360^0$ of the regained BL law absorption coefficient $\alpha$. Figure 8(a) depicts $\alpha$ for the (001) plane in the energy range $4.5\ eV \leq E_{ph} \leq 6.5\ eV$ and 8(b) depicts $\alpha$ for the (010) plane in the same energy range. The energy range is chosen to demonstrate the absorption anisotropy along different directions on these planes. It is evident from the polar plots that the absorption behavior is indeed dependent on light polarization directions. In the (001) plane, the $\alpha$ along the b-axis [010] direction ($\theta = 90^0$) appears to start rising considerably at a much higher energy compared to the a-axis [100] direction ($\theta = 0^0$) demonstrating clear anisotropy. The $E_{edge}$ along the b-axis is blue shifted compared to that along the a-axis in the (001) plane as expected from previous experimental studies [1,2,11]. On the other hand, in the (010) plane, the $\alpha$ along the $X_a$ dipole moment orientation direction ($\theta \cong 25^0$) and the $X_c$ dipole moment orientation direction ($\theta \cong 115^0$) appears to have clear anisotropy at any given energy, although the anisotropy is comparatively smaller than for the (001) plane.

From this discussion, it is clear that the absorption coefficients $(\alpha, \alpha^*)$ can be introduced for different light polarization angles $\theta$ for (001) $\beta - Ga_2O_3$ in energy range $4.9\ eV < E_{ph} \leq 8.65\ eV$ and (010) $\beta - Ga_2O_3$ in energy range $4.65\ eV < E_{ph} \leq 8.65\ eV$. The transmission and decay of photon flux follows the isotropic BL law and provides the generalized absorption coefficient $\alpha$ or needs another weakened absorption coefficient $\alpha^*$ beyond a critical thickness $z_c$ to approximate the phenomena taking place in $\beta - Ga_2O_3$. For photons with $E_{edge} < E_{ph} \leq$ $E_{ph}^0$, $\alpha$ cannot be defined reliably. These observations confirm that the introduction of penetration depth- dependent absorption coefficients can indeed approximate the decay of photon flux in $\beta - Ga_2O_3$ near the excitonic transition energies. But it is always better to revert to the full numerical solution if accurate modeling of the photo response in $\beta - Ga_2O_3$ is required.

## V. POLARIZATION-DEPENDENT PHOTOCURRENT SPECTRA

A functional material must display a detection cutoff energy of $E_{ph} > 4.43\ eV$ to be considered as useful for engineering solar blind photodetectors [5,30]. $\beta - Ga_2O_3$ is of great interest for this purpose [8,31–33]. To evaluate the effectiveness of utilizing $\beta - Ga_2O_3$ for solar blind UV photodetection applications, one must calculate the photoresponsivity $\mathrm{I}_{PR}$ along different polarization directions of this anisotropic material. Eq. (2) can be utilized to obtain the raw $\mathrm{I}_{PR}$ characteristics from the anisotropic dielectric tensor of $\beta - Ga_2O_3$. The numerically calculated $\mathrm{I}_{PR}$ from a $\beta - Ga_2O_3$ slab depth of $z = 1\ \mu m$ for polarization angles in the range $0^0 \leq \theta \leq 360^0$ and photon energies in the range $4.5\ eV \leq E_{ph} \leq 5.5\ eV$ are plotted in log scale in Figures 9(a)-9(b) for (001) and (010) planes respectively. In the plotted energy range as photon energy increases $\mathrm{I}_{PR}$ increases in amplitude for all polarization directions regardless of the plane being observed. Still clear anisotropy is evident for (001) $\beta - Ga_2O_3$ since the $\mathrm{I}_{PR}$ amplitude is ~100 times lower along [010] direction compared to any other directions in the plane for $4.9\ eV > E_{ph} > 4.5\ eV$. It is then expected that $\mathrm{I}_{PR}$ should be minuscule for $E_{ph} < 4.43\ eV$ along b-axis [010] direction. Such a large amplitude anisotropy between the major a- and c-axes, i.e., [100] and [001] directions is not detected in (010) $\beta - Ga_2O_3$, although a small anisotropy between the exciton dipole moment $X_a$ and $X_c$ orientation directions can be seen. Thus, it is evident that the photoresponsivity in the solar irradiation and beyond (up to $\sim 4.9\ eV$) energy range is much more suppressed for the b-axis [010] direction in $\beta - Ga_2O_3$ compared to other major axes, i.e., a- [100] and c- [001] directions.

The $\mathrm{I}_{PR}$ spectra for major axes a- [100] and b- [010] directions in $\beta - Ga_2O_3$ are numerically calculated and shown in Figures 10(a) and 10(b) for depths of $z = 0.4\ \mu m$ and $2\ \mu m$. Experimental measurements on a (001) $\beta - Ga_2O_3$ Schottky diode under proper biasing conditions (0V and 40V, respectively) to produce similar depletion widths are carried out and shown in Figures 10(c) 0V and 10(d) 40V. Details of the device and the measurement procedure can be obtained in our previous works [20,28,34]. The measurements show the same behavior as the calculations, i.e., higher responsivity along [100] direction compared to [010] direction in the $4.5\ eV < E_{ph} < 4.9\ eV$ energy range. This validates our numerical model and points to the fact that (010) $\beta - Ga_2O_3$ photodiodes are not perfectly solar-blind. For (100) or (001) $\beta - Ga_2O_3$ to work as a true solar-blind photodetector, a linear polarizer aligned along [010] direction must be placed in between the light source and the photodiode to have high UV sensitivity and efficiency. This conclusion is also supported by the recent experimental results of Chen et al. [10].

It should also be noted that while experimental data shows similar $I_{PR}$ magnitudes for both [100] and [010] directions in the range $E_{ph}{\sim}E_{edge}$, this is not the case for the computed results. The calculated $I_{PR}$ along [100] direction is much higher than that along [010] direction. To understand this phenomenon, we note that numerical computation of $I_{PR}$ depends on the dielectric tensor of $\beta - Ga_2O_3$ obtained by ellipsometry measurement. Ellipsometry was done on a real $\beta - Ga_2O_3$ single crystalline sample with defects. Such defects will facilitate absorption of sub-bandgap photons and generate immobile photocarriers. Since the photocurrent measurement of the Schottky diode will only collect and quantify mobile charge carriers, it is expected that these locally trapped immobile carriers will not show up in the experimental data. This deviation between simulation and experiment can be utilized to identify localized defect transitions that do not contribute to mobile photocarriers.

## VI. CONCLUSION

In this work, we have solved the EM wave equation following the Berreman-Schubert theory and using the measured dielectric tensor of monoclinic $\beta - Ga_2O_3$ to model the anisotropic absorption process and calculate the photoresponsivity spectra for different polarization angles of incident light on (001) and (010) oriented $\beta - Ga_2O_3$. The strongly anisotropic absorption process in $\beta - Ga_2O_3$ can not be modeled using a single isotropic absorption coefficient. In the vicinity of the absorption edge where strong excitonic peaks are observed, two energy, critical thickness, and polarization state dependent effective absorption coefficients are needed to capture the photon flux decay with penetration depth; whereas at energies well above the absorption edge the isotropic BL law is maintained. The strong linear dichroism property of $\beta - Ga_2O_3$ observed in the examined energy range is the root cause of such behavior. The EM wave equation solution based numerical model predicts that for $E_{ph} < 4.9\ eV$ photoresponsivity is highly anisotropic. Selective solar blind responsivity is predicted for light polarized along the [010] versus [100] directions of $\beta - Ga_2O_3$. These numerical methods can be extended to the entire class of low symmetry wide bandgap semiconductors to enable prediction of photoresponsivity spectra using their measured or predicted dielectric functions. These studies can enable a comprehensive understanding of the optoelectronic properties of low dimensional semiconductors by validating *ab initio* band structure calculations through quantitative comparison of anisotropic photocurrent spectra with ellipsometry measurements.

# Acknowledgements

The work carried out at The Ohio State University was supported by the Center for Emergent Materials, an NSF MRSEC, under award number DMR-2011876. M.S. acknowledges support by the National Science Foundation under awards ECCS 2329940 and OIA-2044049 Emergent Quantum Materials and Technologies (EQUATE), by Air Force Office of Scientific Research under awards FA9550-19-S-0003, FA9550-21-1-0259, and FA9550-23-1-0574 DEF, and by the University of Nebraska Foundation. M.S. also acknowledge support from the J.~A.~Woollam Foundation.

# Appendix A. Electromagnetic Modeling

For solving the complete EM wave equation for the incoming light from air into a $\beta - Ga_2O_3$ slab, the total system, Fig. 1, consists of medium I (air) and medium II ($\beta - Ga_2O_3$). The incident EM wave is travelling parallel to z and the I/II interface is in the x-y plane at $z = 0$. Since the in-plane components of $\vec{E}$ and $\vec{H}$ are conserved, they are related across the interface by,

$$\begin{gathered} \overrightarrow{E_I^+} + \overrightarrow{E_I^-} = \overrightarrow{E_{II}^+} + \overrightarrow{E_{II}^-}, \\ \overrightarrow{H_I^+} + \overrightarrow{H_I^-} = \overrightarrow{H_{II}^+} + \overrightarrow{H_{II}^-}, \end{gathered} \qquad (A1)$$

where $\overrightarrow{E_{I/II}^{+/-}}$ and $\overrightarrow{H_{I/II}^{+/-}}$ represent the forward (+) or backward (-) traveling $\vec{E}$ and $\vec{H}$ components of the EM wave in medium I or II. The electric field can be written as, $\vec{E} = B\overrightarrow{E_1} + C\overrightarrow{E_2}$, where $\overrightarrow{E_1}$ and $\overrightarrow{E_2}$ are its two eigen modes. Assuming there is no back reflection in $\beta - Ga_2O_3$ ($\overrightarrow{E_{II}^-} = \overrightarrow{H_{II}^-} = 0$) then from equation (A2),

$$\begin{gathered} B_1\overrightarrow{E_{I,1}^+} + B_2\overrightarrow{E_{I,2}^+} + C_1\overrightarrow{E_{I,1}^-} + C_2\overrightarrow{E_{I,2}^-} = D_1\overrightarrow{E_{II,1}^+} + D_2\overrightarrow{E_{II,2}^+}, \\ B_1\overrightarrow{H_{I,1}^+} + B_2\overrightarrow{H_{I,2}^+} + C_1\overrightarrow{H_{I,1}^-} + C_2\overrightarrow{H_{I,2}^-} = D_1\overrightarrow{H_{II,1}^+} + D_2\overrightarrow{H_{II,2}^+}. \end{gathered} \qquad (A2)$$

Expanding in terms of their separate x and y components this can be written as,

$$L * \Psi_I = P * \Psi_{II} . \qquad (A3)$$

Such that,

$$L = \begin{bmatrix} \overrightarrow{E_{I,1,x}^+} & \overrightarrow{E_{I,2,x}^+} & \overrightarrow{E_{I,1,x}^-} & \overrightarrow{E_{I,2,x}^-} \\ \overrightarrow{H_{I,1,y}^+} & \overrightarrow{H_{I,2,y}^+} & \overrightarrow{H_{I,1,y}^-} & \overrightarrow{H_{I,2,x}^-} \\ \overrightarrow{E_{I,1,x}^+} & \overrightarrow{E_{I,2,x}^+} & \overrightarrow{E_{I,1,x}^-} & \overrightarrow{E_{I,2,x}^-} \\ \overrightarrow{H_{I,1,y}^+} & \overrightarrow{H_{I,2,y}^+} & \overrightarrow{H_{I,1,y}^-} & \overrightarrow{H_{I,2,x}^-} \end{bmatrix}; P = \begin{bmatrix} \overrightarrow{E_{II,1,x}^+} & \overrightarrow{E_{II,2,x}^+} \\ \overrightarrow{H_{II,1,y}^+} & \overrightarrow{H_{II,2,x}^+} \\ \overrightarrow{E_{II,1,x}^+} & \overrightarrow{E_{II,2,x}^+} \\ \overrightarrow{H_{II,1,y}^+} & \overrightarrow{H_{II,2,x}^+} \end{bmatrix};$$

$$\Psi_I = [B_1 \quad B_2 \quad C_1 \quad C_2]^T; \; \Psi_{II} = [D_1 \quad D_2]^T .$$

Taking, $\mu = L^{-1} * P$, the Berreman wave equation [25] is written as,

$$\Psi_I = L^{-1} * P * \Psi_{II} = \mu * \Psi_{II} , \qquad (A4)$$

Additionally,

$$\Psi_I = \left[\overrightarrow{I_a}, \overrightarrow{R_a}\right]^T; \; \Psi_{II} = \left[\overrightarrow{T_a}\right]^T, \qquad (A5),$$

where $\overrightarrow{I_a}$ is the amplitude vector of the incident electric/magnetic field, $\overrightarrow{R_a}$ is the amplitude vector of the reflected electric/magnetic field, and $\overrightarrow{T_a}$ is the amplitude vector of the transmitted electric/magnetic field. Also, $\mu = \begin{bmatrix} t^{-1} \\ \eta \end{bmatrix}$. Then,

$$\overrightarrow{I_a} = t^{-1} * \overrightarrow{T_a}; \; \overrightarrow{T_a} = t * \overrightarrow{I_a},$$
$$\overrightarrow{R_a} = \eta * \overrightarrow{T_a} = \eta * t * \overrightarrow{I_a} = r * \overrightarrow{I_a}. \qquad (A6)$$

Here, $t$ corresponds to transmission matrix and $r$ corresponds to reflection matrix. To find the Jones matrices t and $r$ we utilize the theory of Berreman [25] which has been modified by Schubert [26]. The method has high similarity to the works of Lin-chung [35] and Ong [36]. The method requires forming a Delta matrix which requires the complete description of the dielectric tensor including the off-diagonal components. The complete dielectric tensor of $\beta - Ga_2O_3$ in the energy range $E_{ph} = 2 - 8.65\ eV$ had been obtained via ellipsometry by Sturm et al. [17] and is utilized for all subsequent calculations in this work.

A physical polarizer set up on light's path varies the polarization angle $\theta$ leading to changes in the incident light field components, $\overrightarrow{E_\iota}$ and $\overrightarrow{H_\iota}$, i.e.,

$$\overrightarrow{E_\iota} = \begin{bmatrix} cos\theta & -sin\theta \\ sin\theta & cos\theta \end{bmatrix} * \begin{bmatrix} \vec{E}_x \\ \vec{E}_y \end{bmatrix}, \qquad (A7)$$

where $\theta = 0$ corresponds to light polarized along the x-axis. The transmitted and reflected light's E-fields are given, respectively, as,

$$\overrightarrow{E_t} = t * \overrightarrow{E_\iota}; \; \overrightarrow{E_r} = r * \overrightarrow{E_\iota}. \qquad (A8)$$

For $\overrightarrow{E_t}$, an explicit z-dependence is calculated in order to model the absorption process. Analogous equations (A7 and A8) are used to compute $\overrightarrow{H_t}$ and $\overrightarrow{H_r}$.

The angle of incidence ($\varphi$) with respect to the surface normal ($-z$) gives the k-vector $\vec{k} = \vec{k}_x + \vec{k}_y$ components defined as $k_x = \sin(\varphi)$, $k_y = \cos(\varphi)$. It should be noted that we choose normal incidence on slab case for our calculations i.e., $\varphi = 0^0$ to simplify the problem. The field magnitudes of reflected light are calculated by,

$$E_r = E_{pr} * E_x + E_{sr} * E_y; \; H_r = H_{pr} * E_x + H_{sr} * E_y, \qquad (A9)$$

Here, $E_{pr} = \left[k_y \; 0 \; k_x\right], E_{sr} = [0\;1\;0], H_{pr} = [0 - 1\;0], H_{sr} = [-k_x\;0 - k_y]$.

Similarly, the incident light field magnitudes are,

$$E_i = E_{pi} * E_x + E_{si} * E_y;\ H_i = H_{pi} * E_x + H_{si} * E_y\ . \qquad (A10)$$

Here, $E_{pi} = [k_y\ 0\ -k_x], E_{si} = [0 - 1\ 0], H_{pi} = [0\ 1\ 0], H_{si} = [k_y\ 0\ k_x]$.

The incoming light's time averaged Poynting vector before surface reflection or absorption is [37],

$$\overrightarrow{S_{in}} = \frac{1}{4}\left(\overrightarrow{E_i}\ X\ \overrightarrow{H_i^*} + \overrightarrow{E_i^*}\ X\ \overrightarrow{H_i}\right), \qquad (A11)$$

with $\overrightarrow{E_i} = \vec{E}_x + \vec{E}_y$ and $\overrightarrow{H_i} = \vec{H}_x + \vec{H}_y$; $\overrightarrow{E_i^*}$ and $\overrightarrow{H_i^*}$ are their conjugates.

Similarly, the transmitted light's $(z > 0)$ time averaged Poynting vector into $\beta - Ga_2O_3$ is given as,

$$\overrightarrow{S_t} = \frac{1}{4}\left(\overrightarrow{E_t}\ X\ \overrightarrow{H_t^*} + \overrightarrow{E_t^*}\ X\ \overrightarrow{H_t}\right). \qquad (A12)$$

And the reflected light's time averaged Poynting vector is,

$$\overrightarrow{S_r} = \frac{1}{4}\left(\overrightarrow{E_r}\ X\ \overrightarrow{H_r^*} + \overrightarrow{E_r^*}\ X\ \overrightarrow{H_r}\ \right). \qquad (A12)$$

The photon flux transmitting into $\beta - Ga_2O_3$ is,

$$\Phi = \frac{\left|\overrightarrow{S_t}\right|}{E_{ph}}\ . \qquad (A13)$$

This transmitted photon flux gets absorbed in the slab and produces photocarriers as described by $G$ given in equation (1) in the main text which is numerically calculated from the three-dimensional $\left|\overrightarrow{S_t}\right|$ data, i.e., dependent on $E_{ph}, \theta$, and $z$. The photocurrent density $J$ is calculated by integrating $G$ to a limit $z = d$,

$$J = q\int_0^d G\ dz\ , \qquad (A14)$$

where q is the electron charge and $d$ is the width of the photocarrier collection region, assuming a Schottky diode with depletion region beginning at $z = 0$. The photoresponsivity $I_{PR}$ is then given as,

$$\mathrm{I}_{PR} = \frac{\mathrm{J}}{\left|\overrightarrow{S_{in}}\right|}. \qquad (A15)$$

Equation (A15) which is equivalent to equation (2) in main text is the purely numerical definition of photoresponsivity of $\beta - Ga_2O_3$. Depth $z$ can be varied over different $\beta - Ga_2O_3$ collection width $d$ to obtain different $I_{PR}$.

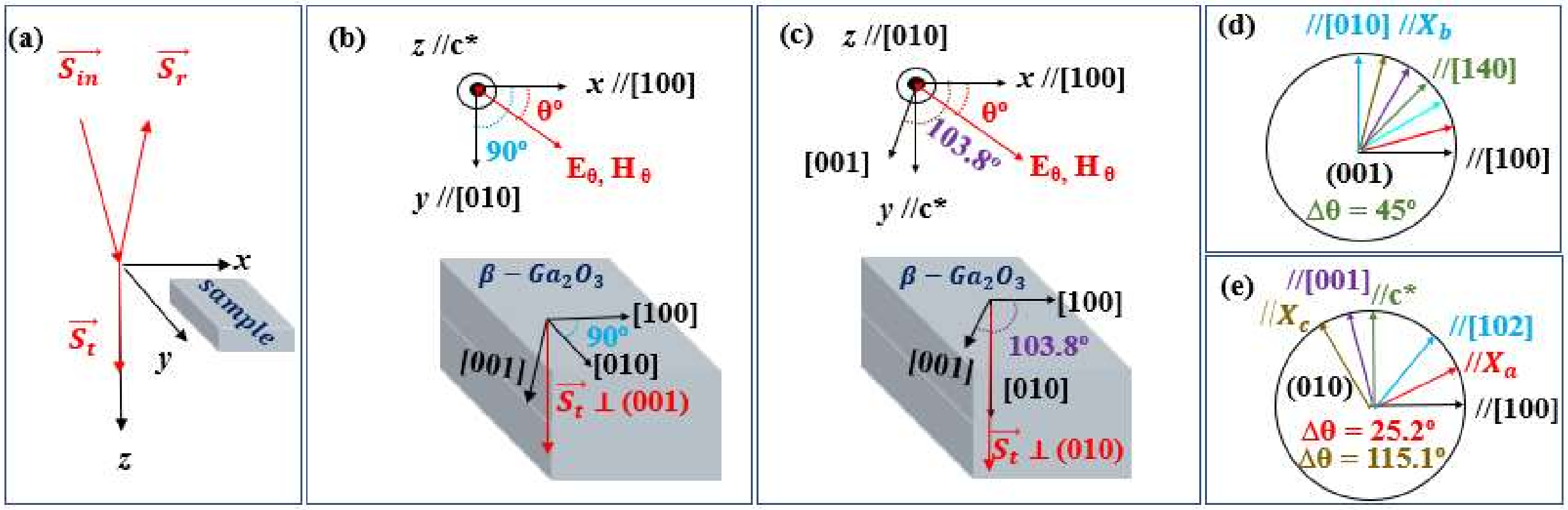

Figure 1: (a) Computational setup for EM wave propagation modeling into $\beta - Ga_2O_3$ sample sitting in the lab frame, (b) Top view and side view of (001) oriented $\beta - Ga_2O_3$ sample; different directions are shown with respect to the lab frame, (c) Top view and side view of (010) oriented $\beta - Ga_2O_3$ sample; different directions are shown with respect to the lab frame, (d) Polarization directions of incident light on (001) plane of $\beta - Ga_2O_3$ along with exciton dipole

moment orientation of $X_b$ and, (e) Polarization directions of incident light on (010) plane of $\beta - Ga_2O_3$ along with exciton dipole moment orientations of $X_a$ and $X_c$.

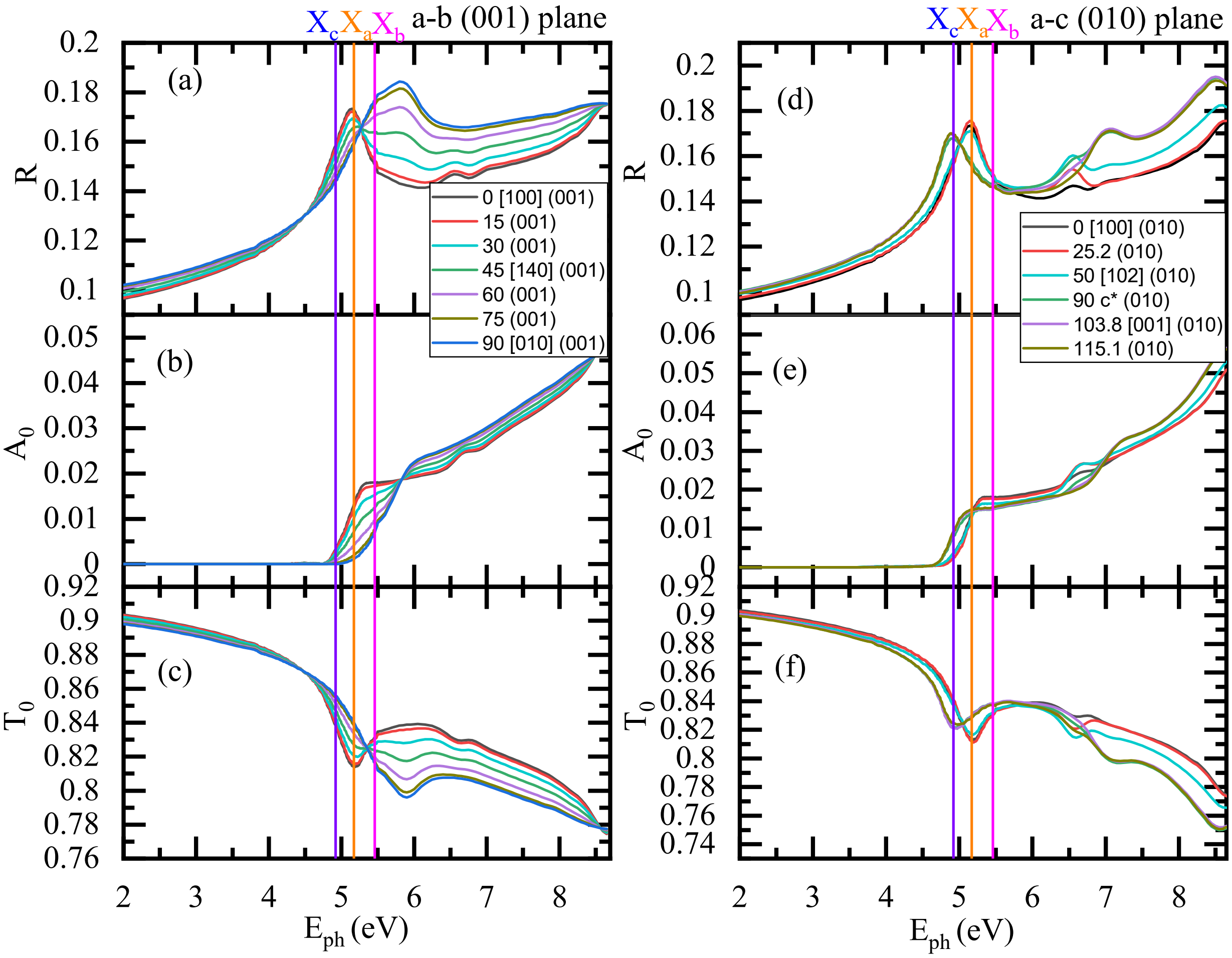

Figure 2: The light polarization dependent (a) surface $(z = 0)$ reflectance spectra from the Air/(001) $\beta - Ga_2O_3$ interface (b) absorbance and (c) transmittance spectra inside the surface $(z = 1\,nm)$ of (001) $\beta - Ga_2O_3$ for polarization angles (legend in degrees) $0^0 \leq \theta \leq 90^0$ at $\Delta\theta = 15^0$ steps in the energy range $2\,eV \leq E_{ph} \leq 8.65\,eV$. The light polarization dependent (d) surface $(z = 0)$ reflectance spectra from the Air/(010) $\beta - Ga_2O_3$ interface (e) absorbance and (f) transmittance spectra inside the surface $(z = 1\,nm)$ of (010) $\beta - Ga_2O_3$ for polarization angles (legend in degrees) $0^0 \leq \theta \leq 180^0$ at $\theta = 0^0, 25.2^0, 50^0, 90^0, 103.8^0, 115.1^0$ in the energy range $2\,eV \leq E_{ph} \leq 8.65\,eV$. Excitonic transition peaks' energies are denoted by $X_c, X_a$ and $X_b$ in all subfigures.

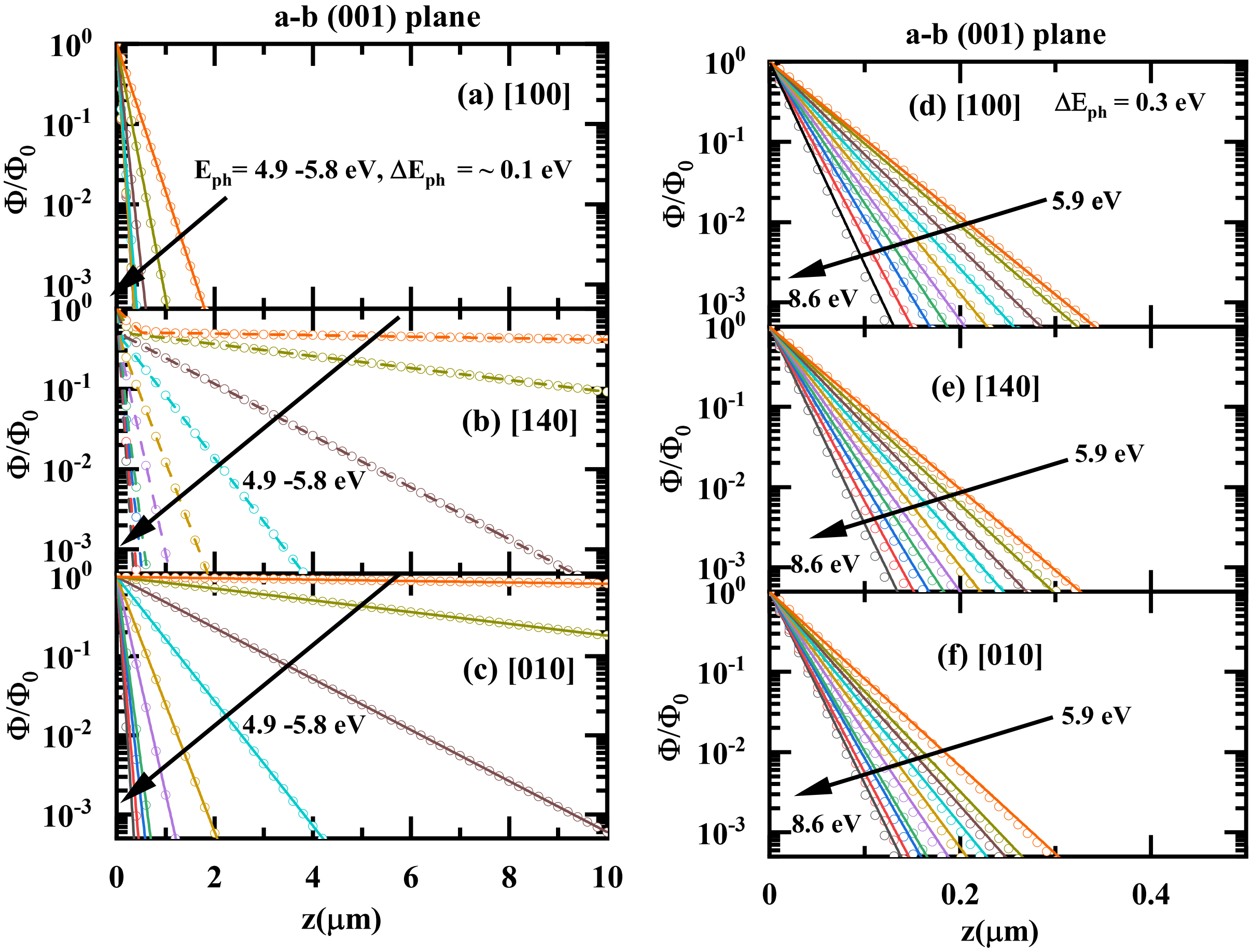

Figure 3: The flux decay into (001) $\beta - Ga_2O_3$ shown at specific energies at $\cong\ 0.1\,eV$ steps with an arrow for increasing values near absorption edge in the range $4.9\,eV < E_{ph} \leq 5.8\,eV$ for polarization angles (a) $\theta = 0^0$ [100] direction, (b) $\theta = 45^0$ [140] direction, (c) $\theta = 90^0$ [010] direction in the (001) plane. The BL law fits are shown in solid lines for [100] direction and [010] direction while two exponential decay fits are shown in dashed lines for [140] direction. The flux decay into (001) $\beta - Ga_2O_3$ shown at specific energies at $0.3\,eV$ steps with an arrow for increasing values well above the absorption edge in the range $5.8\,eV < E_{ph} \leq 8.65\,eV$ for polarization angles (d) $\theta = 0^0$ [100] direction, (e) $\theta = 45^0$ [140] direction, (f) $\theta = 90^0$ [010] direction in the (001) plane. The BL law fits are shown using solid lines for all directions.

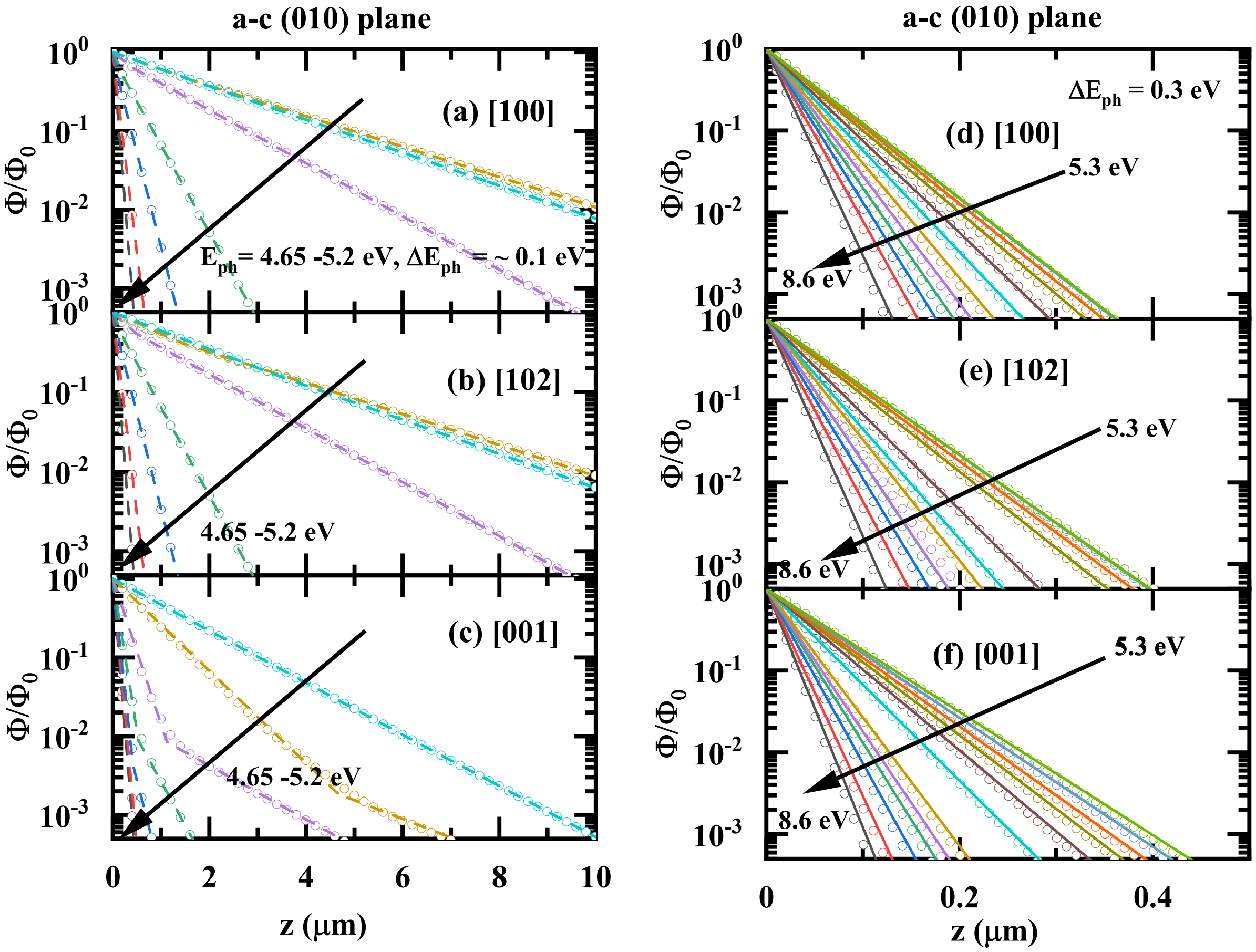

Figure 4: The flux decay into (010) $\beta - Ga_2O_3$ at specific energies at $\cong 0.1\ eV$ steps shown with an arrow for increasing values near absorption edge in the range $4.65\ eV < E_{ph} \leq 5.20\ eV$ for polarization angles (a) $\theta = 0^0$ [100] direction, (b) $\theta = 50^0$ [102] direction, (c) $\theta = 103.8^0$ [001] direction in the (010) plane. The two exponential decay fits are shown in dashed lines for all directions. The flux decay into (010) $\beta - Ga_2O_3$ at specific energies at $0.3\ eV$ steps shown with an arrow for increasing values well above the absorption edge in the range $5.2\ eV < E_{ph} \leq 8.65\ eV$ for polarization angles (d) $\theta = 0^0$ [100] direction, (e) $\theta = 50^0$ [102] direction, (f) $\theta = 103.8^0$ [001] direction in the (010) plane. The BL law fits are shown using solid lines for all directions.

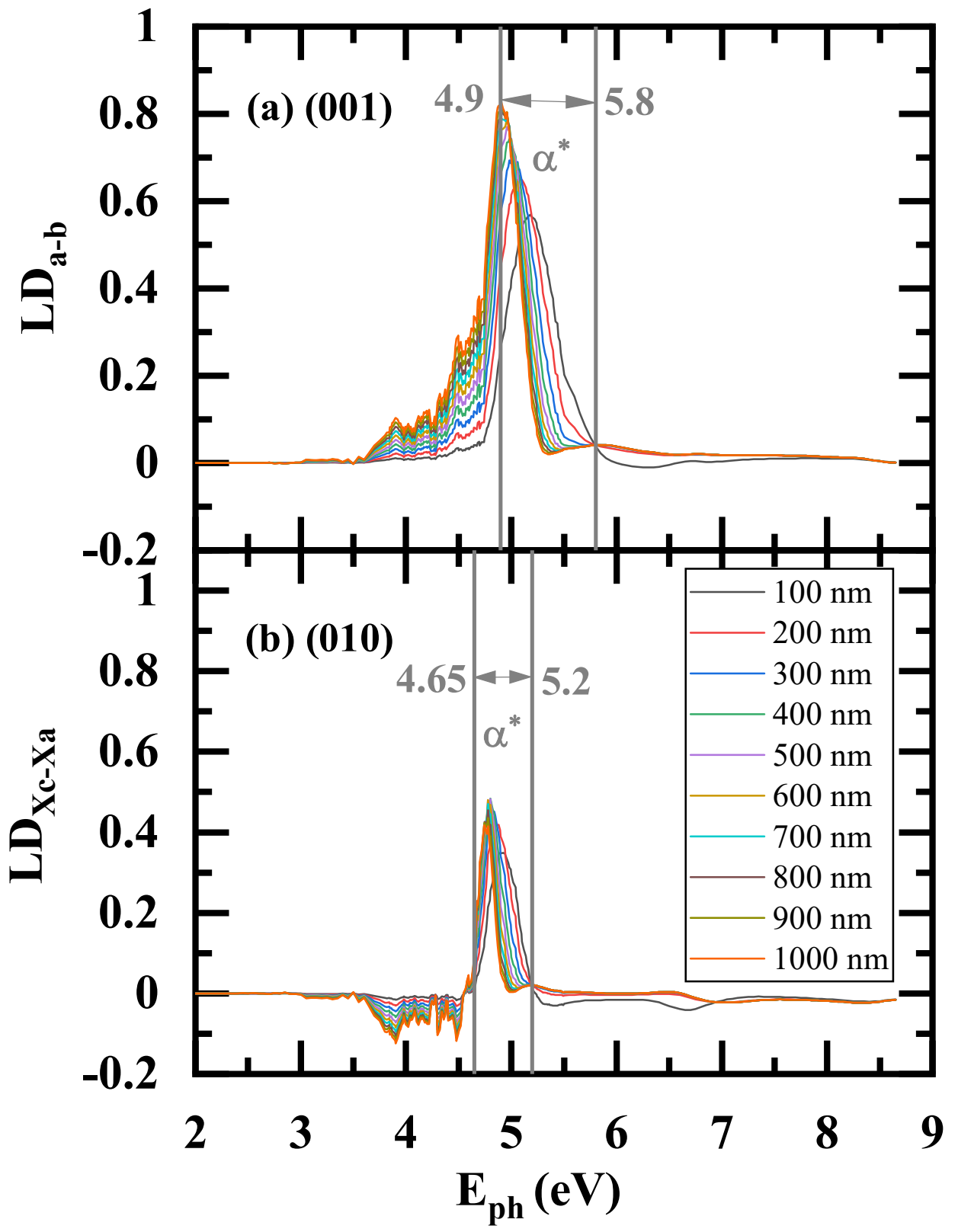

Figure 5: The Linear Dichroism (LD) spectra for light travelling perpendicular to (a) (001) plane and (b) (010) plane of $\beta - Ga_2O_3$ at different thicknesses $z \leq 1\ \mu m$ obtained by subtracting the absorbance along two perpendicular directions in the plane i.e., a-axis - b-axis in the (001) plane and $X_c$ direction - $X_a$ direction in the (010) plane. Energy range of weakened absorption coefficient $\alpha^*(/cm)$ overlaps with large LD expectations.

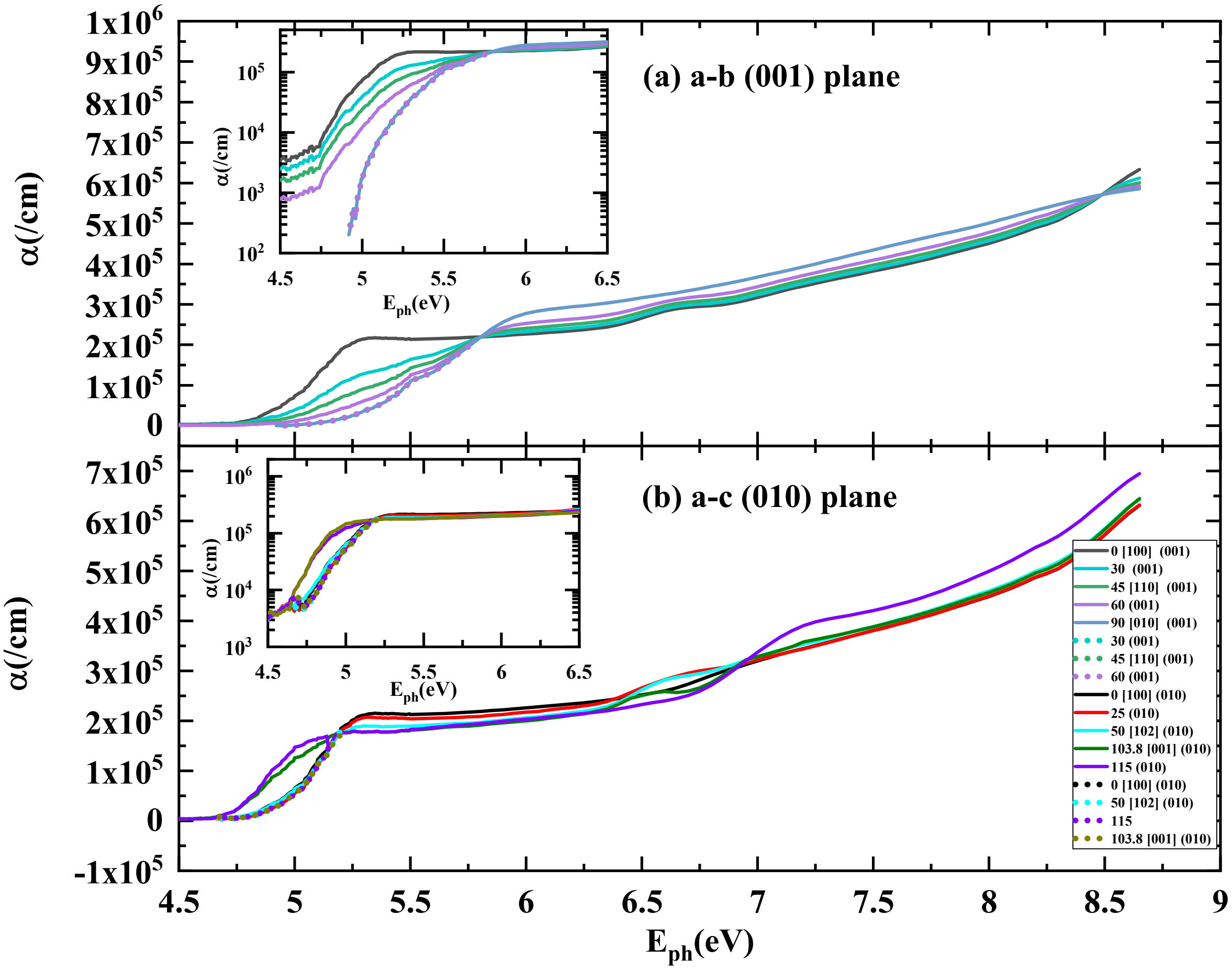

Figure 6: (a) The generalized absorption coefficient $\alpha(/cm)$ (solid lines) extracted from BL law and weakened absorption coefficient $\alpha^*(/cm)$ (dots) extracted from two absorption coefficient decay model for energies in the range $4.5\,eV \leq E_{ph} \leq 8.65\,eV$ for specific directions within polarization angles $0^0 \leq \theta \leq 90^0$ for (001) $\beta - Ga_2O_3$. The inset shows the results in log scale for energies $4.5\,eV \leq E_{ph} \leq 6.5\,eV$. The weakened $\alpha^*(/cm)$ for all polarizations shown follows the generalized $\alpha(/cm)$ along b-axis i.e., $X_b$ exciton dipole moment's orientation. (b) The generalized absorption coefficient $\alpha(/cm)$ (solid lines) extracted from BL law and weakened absorption coefficient $\alpha^*(/cm)$ (dots) extracted from two absorption coefficient decay model for energies in the range $4.5\,eV \leq E_{ph} \leq 8.65\,eV$ for specific directions within polarization angles $0^0 \leq \theta \leq 180^0$ for (010) $\beta - Ga_2O_3$. The inset shows the results in log scale for energies $4.5\,eV \leq E_{ph} \leq 6.5\,eV$. The weakened $\alpha^*(/cm)$ for all polarizations shown follows the generalized $\alpha(/cm)$ along $X_a$ exciton dipole moment's orientation.

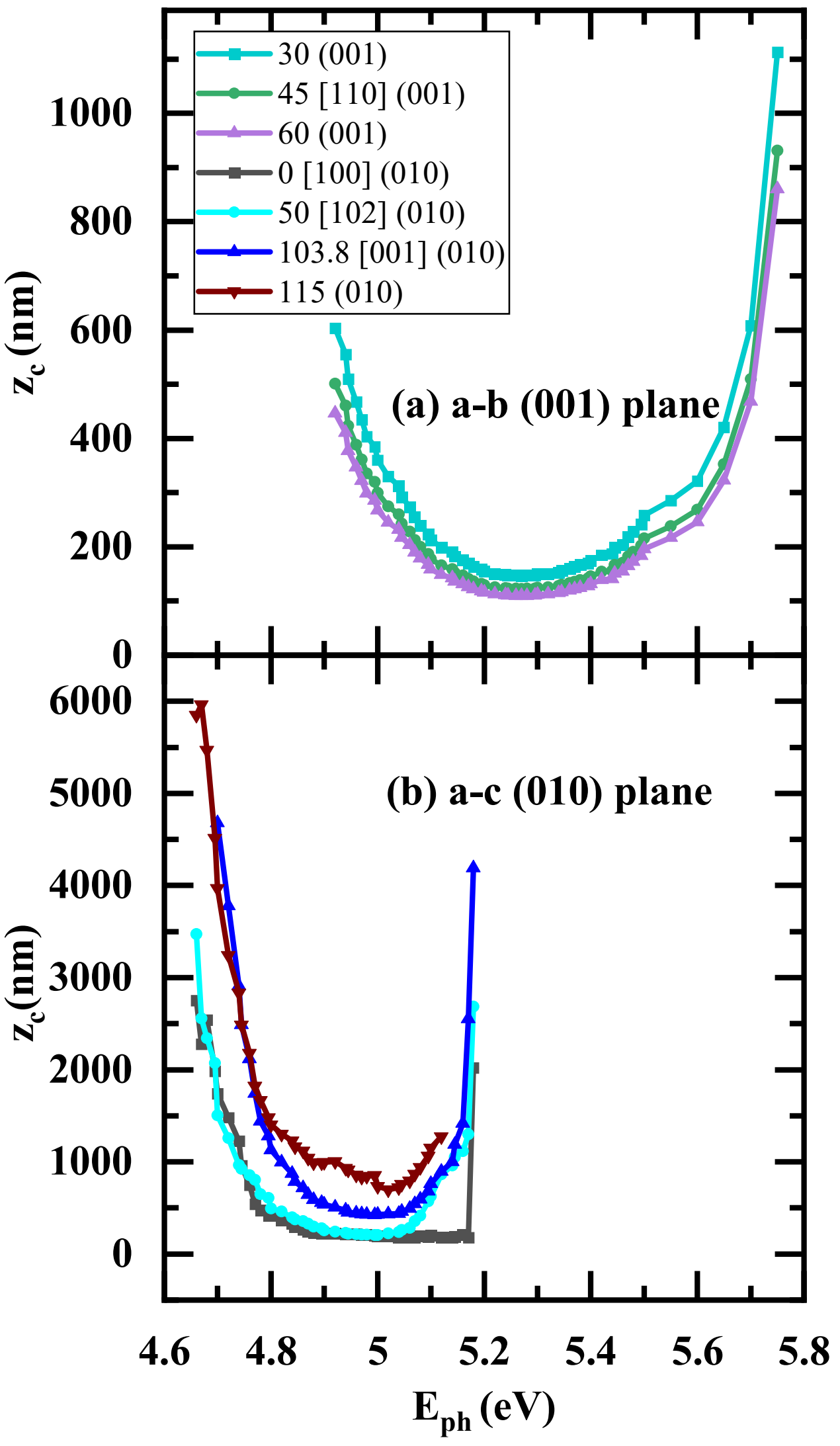

Figure 7: (a) The critical penetration depth $z_c\ (nm)$ extracted from two absorption coefficients decay model for energies in the range $4.9\ eV < E_{ph} \leq 5.8\ eV$ for specific directions within polarization angles $0^0 < \theta < 75^0$ in (001) $\beta - Ga_2O_3$. (b) The critical penetration depth $z_c\ (nm)$ extracted from two absorption coefficients decay model for energies in the range $4.65\ eV < E_{ph} \leq 5.2\ eV$ for specific directions within polarization angles $0^0 \leq \theta \leq 140^0$ in (010) $\beta - Ga_2O_3$. The critical penetration depth decreases as the incoming polarization angle aligns closer to the least absorbing polarization direction.

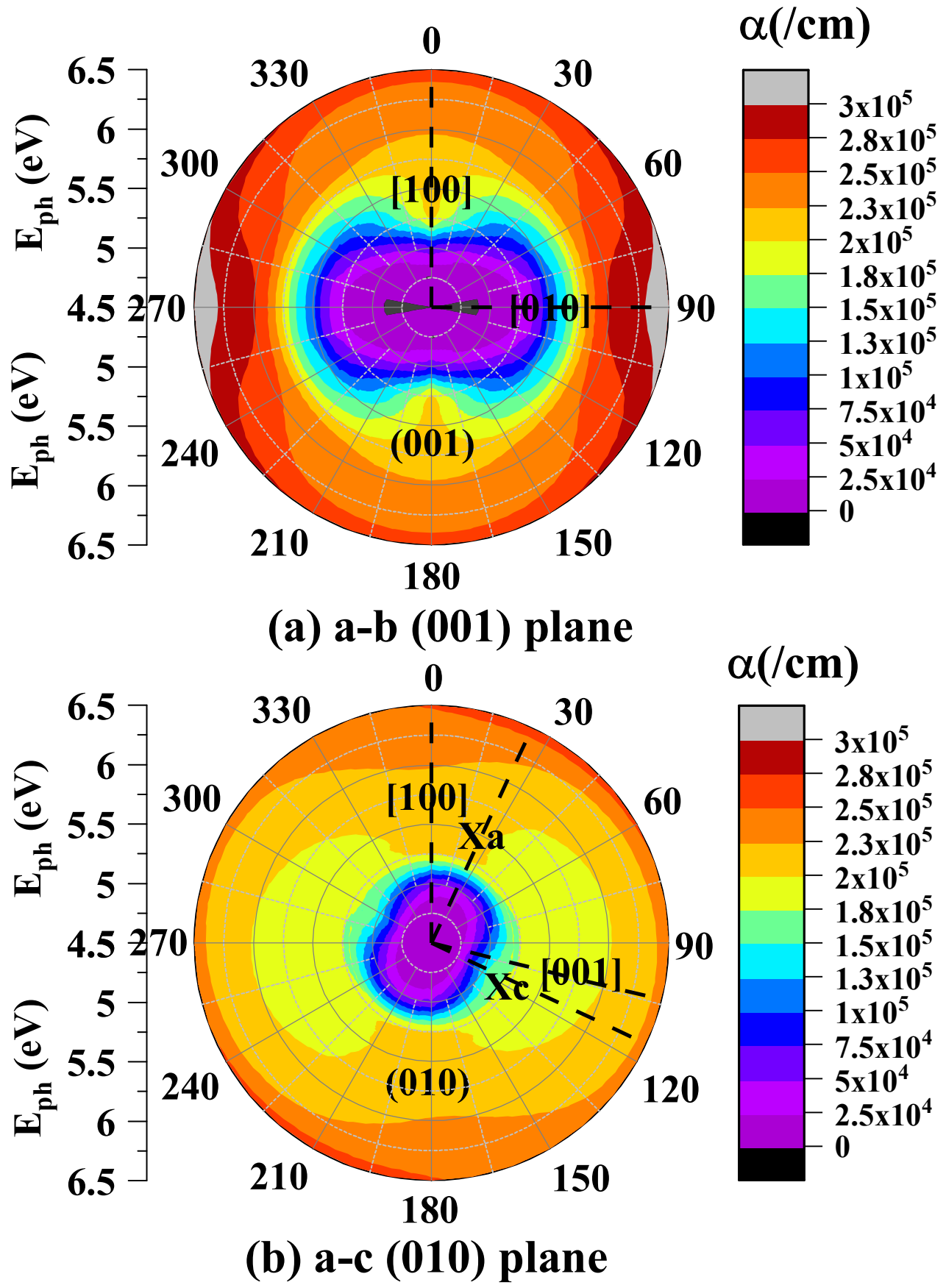

Figure 8: Polar plot of the generalized absorption coefficient $\alpha\ (/cm)$ spectra in the energy range $4.5\ eV \leq E_{ph} \leq 6.5\ eV$ for polarization angles $0^0 \leq \theta \leq 360^0$ at $\Delta\theta = 5^0$ steps for (a) (001) $\beta - Ga_2O_3$ and (b) (010) $\beta - Ga_2O_3$ . The $\alpha\ (/cm)$ anisotropy between a-axis [100] $\theta = 0^0$ direction and b-axis [010] $\theta = 90^0$ direction in the (001) plane is much more pronounced compared to the anisotropy between $X_a$ dipole moment orientation $\theta \cong 25^0$ direction and $X_c$ dipole moment orientation $\theta \cong 115^0$ direction in the (010) plane

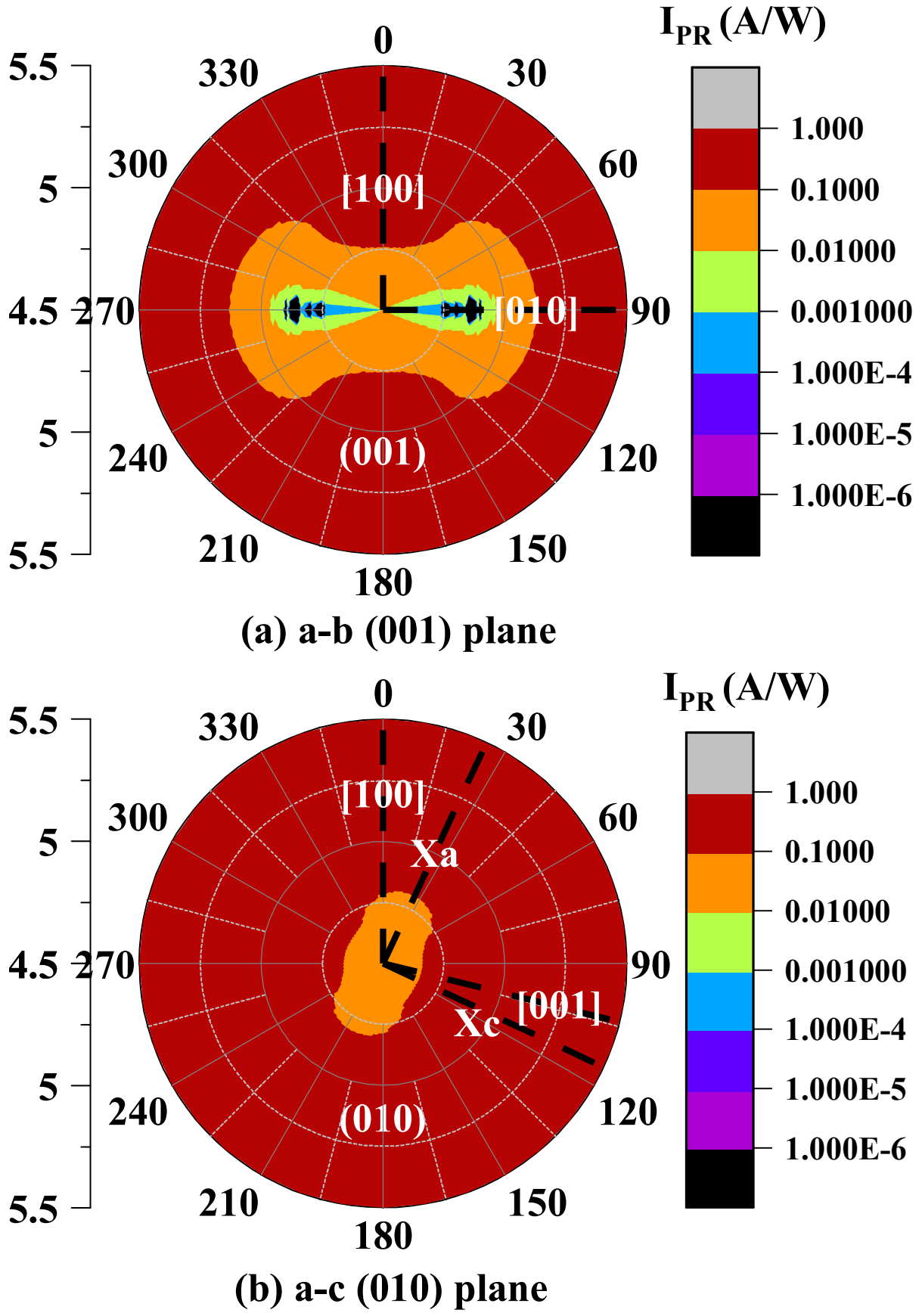

Figure 9: Polar plots of the photoresponsivity $I_{PR}$ $(A/W)$ spectra in log scale in the energy range $4.5\ eV \le E_{ph} \le 5.5\ eV$ for polarization angles $0^0 \le \theta \le 360^0$ at $\Delta\theta = 5^0$ steps for $\beta - Ga_2O_3$ slab thicknesses $z = 1\ \mu m$ for (a) (001) and (b) (010) planes. The $I_{PR}$ $(A/W)$ anisotropy between a-axis [100] $\theta = 0^0$ direction and b-axis [010] $\theta = 90^0$ direction in the (001) plane is much more pronounced compared to the anisotropy between $X_a$ dipole moment orientation $\theta \cong 25^0$ direction and $X_c$ dipole moment orientation $\theta \cong 115^0$ direction in the (010) plane.

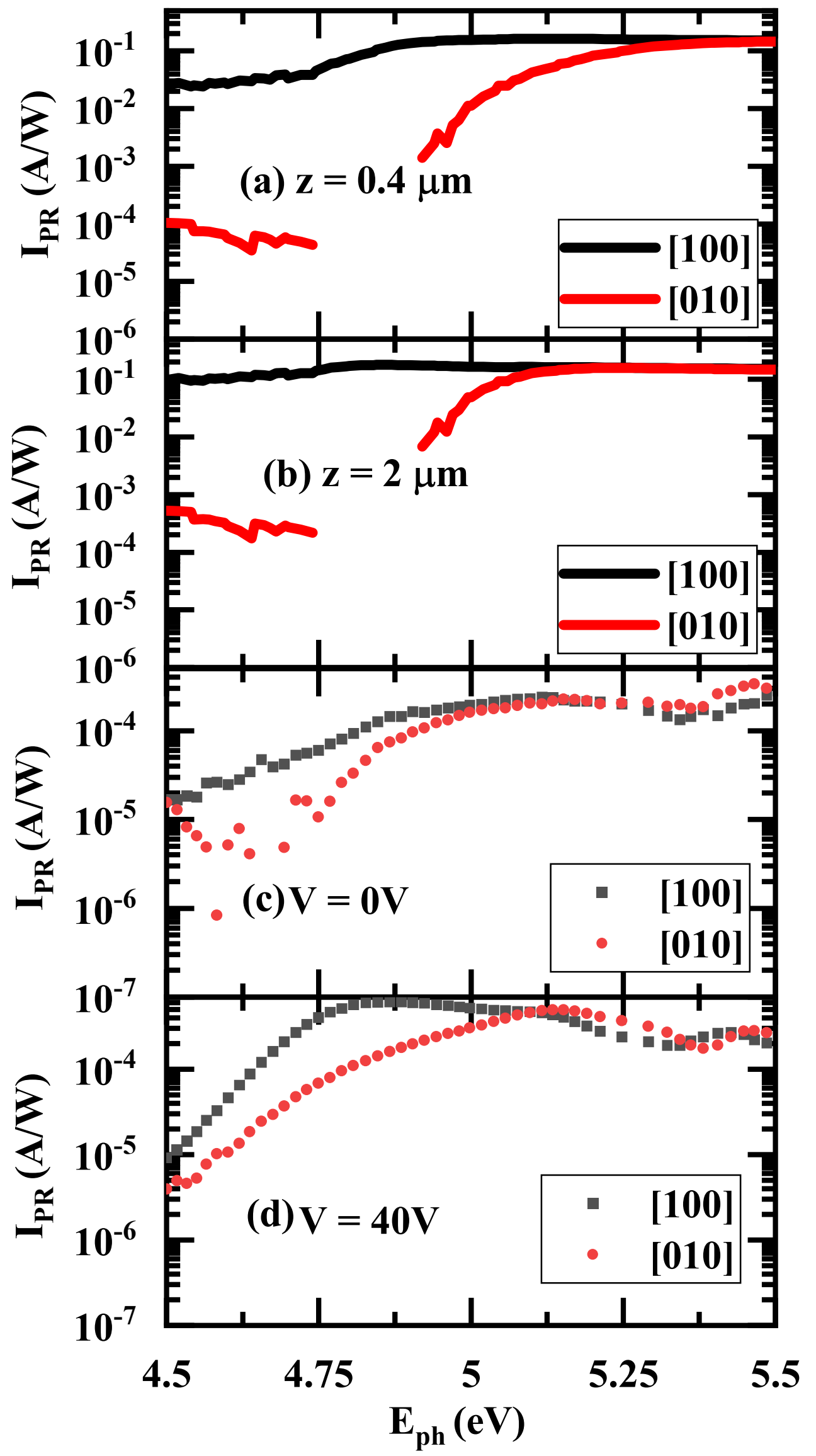

Figure 10: The photoresponsivity $I_{PR}$ $(A/W)$ spectra excited by light polarized along [100] and [010] directions for $\beta - Ga_2O_3$ thicknesses (a) $z = 0.4\ \mu m$, (b) $z = 2\ \mu m$ from simulations in the range $4.5\ eV \leq E_{ph} \leq 5.5\ eV$. These results are compared with (001) $\beta - Ga_2O_3$ Schottky diode measurements done at biases (c) $V = 0V$ and (d) $V = 40V$ at directions [100] and [010]. Both simulations and measurements show suppression of $I_{PR}$ along [010] direction compared to [100] direction below $4.9\ eV$ .